\documentclass[letterpaper,10pt]{article}
\usepackage{opex3}

\usepackage{epsfig}
\usepackage{amsmath,amssymb}

\begin{document}

\title{A Compact Orbital Angular Momentum Spectrometer Using Quantum Zeno Interrogation}

\author{Paul Bierdz, Hui Deng*}
\address{Department of Physics, University of Michigan, Ann Arbor, USA}
\email{paopao@umich.edu, dengh@umich.edu}
\begin{abstract} We present a scheme to measure the orbital angular momentum spectrum of light using a precisely timed optical loop and quantum non-demolition measurements.  We also discuss the influence of imperfect optical components.
\end{abstract}

\bibliographystyle{osajnl}	

Light carries quantized orbital angular momentum (OAM) of $l \hbar$ per photon when the electric field has an overall azimuthal dependence on phase, $e^{-i l \phi}$, where $\phi$ is the azimuthal angle about the beam propagation axis \cite{allen_orbital_1992,molina-terriza_twisted_2007,franke-arnold_advances_2008}.  The OAM quantum number $l$ takes integer values from $-\infty$ to $+\infty$.  With an infinite number of states available, OAM can be utilized for qudit ($d$-dimensional, $d>2$) systems \cite{molina-terriza_management_2001} that allow, for example, higher dimensional entanglement \cite{mair_entanglement_2001}, quantum coin-tossing \cite{molina-terriza_experimental_2005}, increased violations of local realism \cite{kaszlikowski_violations_2000}, improved security for quantum key distribution \cite{cerf_security_2002}, simplified quantum gates \cite{lanyon_simplifying_2009} and superdense coding for quantum communications with increased channel capacity \cite{barreiro_beating_2008}.

Determining OAM of light in an unknown state, however, is more challenging than measuring different polarizations or frequencies.  Eigenstates of OAM can be deduced from the diffraction interference pattern with judiciously chosen apertures \cite{hickmann_unveiling_2010,wang_novel_2009}.  But the method requires a large collection of photons to develop the pattern.  Moreover it may become very complicated for superpositions of OAM states.  An $l$-fold fork diffraction grating \cite{bazhekov_laser-beams_1990,heckenberg_generation_1992,arlt_production_1998} can separate a pre-determined OAM component from others at the single photon level, but cannot be readily applied to determine arbitrary OAM state of light.  A cascade Mach-Zehnder interferometer setup, using Dove prisms to introduce an $l$-dependent phase shift, can separate different OAM components of light into different output ports of the interferometers, even at the single-photon level \cite{leach_interferometric_2004}.  But the setup requires $N-1$ mutually stabilized interferometers to detect $N$ OAM-modes.  A recently proposed scheme \cite{berkhout_efficient_2010} requires only a spatial light modulator (or a specially designed hologram), which converts the twisting phase structure of OAM states into a linear phase gradient, and a lens, which focuses different OAM components to different spatial locations on the focal plane.  However, the method relies on intricate spatial modulation of the phase, and thus may have limited applicability to broadband ultrafast pulses.  Moreover, the extinction ratio between different OAM states is limited to about $10$.  To increase the extinction ratio or to detect higher order OAM states will require larger and increasingly complex holograms.

\begin{figure}[tb]
  \centering
  \includegraphics[width=0.95\textwidth]{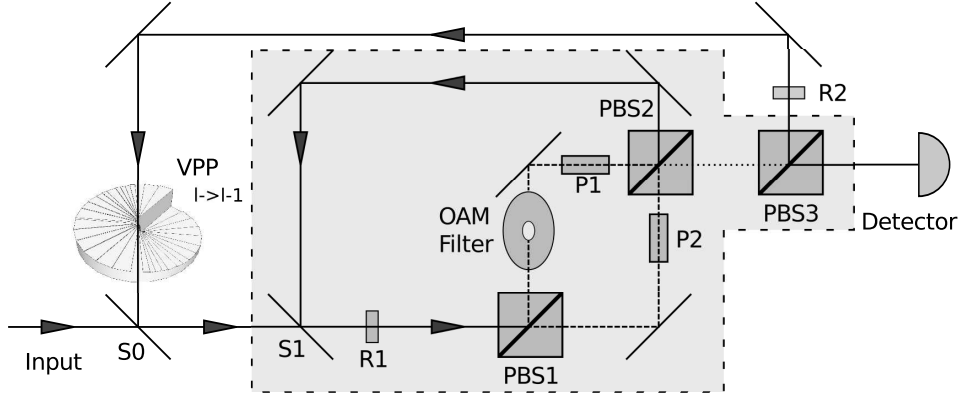}
\caption{A schematic of the compact OAM spectrometer.  The Quantum Zeno Interrogator (shaded region) distinguishes between zero and nonzero OAM states. The outer loop decreases the OAM value of light by one per round trip. All the beam splitters are polarizing beam splitters (PBSs) that transmits horizontally polarized light and reflects vertically polarized light. The OAM filter transmits states with zero OAM, but blocks states with non-zero OAM.  S0 and S1 are switching mirrors that either transmits or reflects incident light \cite{optical_switch}.  R1 and R2 are fixed polarization rotators, which can be half wave plates.  P1 and P2 are fast polarization switches, such as Pockels cells.  When activated, P1 and P2 switches horizontal polarization to vertical and vice versa.  When de-activated, they are transparent to light.  The shaded region is a Quantum Zeno Interrogator \cite{kwiat_high-efficiency_1999} which separates OAM components with $l=0$ and $l\neq 0$ into different polarizations.  Hence at PBS3, zero OAM component is sent to the detector while the none-zero OAM component is sent back into the outer-loop.  The outer loop decreased OAM by one per round trip via, for example, a vortex phase plate (VPP) \cite{beijersbergen_helical-wavefront_1994}.}
\label{fig:setup}
\end{figure}

In this paper, we present a compact OAM-spectrometer comprising of only one interferometer nested within an optical loop (Fig.~\ref{fig:setup}).  It uses a Quantum Zeno Interrogator (QZI) \cite{peres_zeno_1980,elitzur_quantum_1993,kwiat_high-efficiency_1999} (shaded region in Fig.~\ref{fig:setup}) to perform counterfactual measurements on the OAM state, and thus maps different OAM components of an arbitrary input light pulse into different time bins at the output.  It can achieve very high extinction ratios between different OAM states and can work for arbitrarily high OAM orders limited mainly by optical losses.


We illustrate now how the spectrometer works by tracing, as an example, a horizontally polarized input pulse with an OAM value $l=l_0 \geq 0$, noted as $|\psi^{(0)} \rangle = |H,l_0\rangle$.  The input pulse first transmits through optical switches S0 and S1 \cite{optical_switch}, and enters the QZI.  The polarization rotator R1 rotates its polarization by $\Delta\theta = \pi/(2N)$, and the state becomes $|\psi^{(0)}_1\rangle = \cos\left(\frac{\pi}{2N}\right)|H,l_0\rangle + \sin\left(\frac{\pi}{2N}\right) |V,l_0\rangle$. If $l=0$, the horizontal and vertical components of $|\psi^{(0)}_1 \rangle$ passes through the lower and upper arms of the interferometer, respectively.  They recombine into the same state~$|\psi^{(0)}_1\rangle$ at the polarizing beam splitter PBS2 (neglecting an overall phase factor).  S1 is switched to be reflective at the end of the first QZI loop, and the combined beam continues to loop in the QZI.  The polarization is rotated by $\Delta\theta = \pi/(2N)$ each loop.  After $N$ loops, the light becomes vertically polarized and enters only the upper path of the interferometer.  At this point, the polarization switch P1 is activated and switches the polarization into horizontal. Hence the light transmits through both PBS2 and PBS3, and arrives at the detector at time $T_0$.

If $l_0\neq 0$, however, the vertical component is sent to the upper path at PBS1, and is then blocked by the OAM filter.  Only the horizontal component emerges after PBS2, the state collapses into $|H,l_0\rangle $ with a probability $\cos^2\left(\frac{\pi}{2N}\right)$. After $N$ loops, a fraction $p = \cos(\pi/(2N))^{2N}$ of the light remains in the horizontal polarization in the lower arm of the interferometer, while a fraction $1-p$ of the light is lost (blocked by OAM filter).  At this point, the polarization switch P2 is activated and switches the polarization to vertical, and the light reflects off both PBS2 and PBS3, and enters the outer loop.  By this time, S0 is switched to be reflective.  As the light cycles in the outer loop, the polarization in rotated back to horizontal by R2, and the OAM value is decreased by $\Delta l=1$ per cycle by a vortex phase plate (VPP) \cite{beijersbergen_helical-wavefront_1994}.  After $l_0$ cycles, $l=0$.  When the light enters the QZI again, it will exit the spectrometer to the detector, at a time $T(l_0)=T_0 + l_0 (NL_{QZI}+L_{out})/c$.  Here $L_{QZI}$ and $L_{out}$ are the optical path lengths of the QZI loop (from S1 to PBS2 back to S1) and the outer-loop (from S1 to PBS2, to PBS3, to S0, back to S1).  The detected fraction of the light intensity is $P(l_0)=p^{l_0}=\cos\left(\frac{\pi}{2N}\right)^{2Nl_0}$.

\begin{figure}[tb]
  \centering
  \includegraphics[width=.98\textwidth]{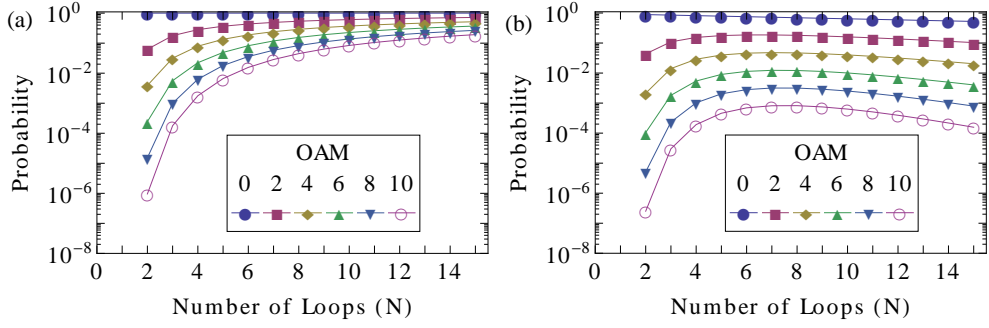}
  \caption{The probability of detecting the correct OAM value as a function of the number of loops ($N$) in the QZI using a perfect OAM filter. (a) Neglect optical loss. (b) Assume $|\alpha|^2 = 0.96$ based on commercially available optics. When optical loss is included, there exists an optimal $N$ for higher order OAM states, due to the compromise between the quantum Zeno enhancement and optical loss.}
  \label{fig:loop_opt_perfectFilter}
\end{figure}

In short, the OAM spectrometer sorts different OAM components into different time intervals separated by $\Delta T=(NL_{QZI}+L_{out})/c$ with a \emph{perfect} extinction ratio. The total transmission efficiency of the spectrometer is $P(l_0)$ for the component with OAM of $l_0\hbar$. $P(l_0)\rightarrow 1$ for all $l_0$ as $N\rightarrow \infty$ due to the quantum Zeno effect \cite{misra_zenos_1977}, as shown in the Fig.~\ref{fig:loop_opt_perfectFilter}(a).

In practice, optical components introduce loss.  Assuming high quality, but commercially available optical components, we estimate a round trip transmission of $|\alpha|^2\sim 0.96$ \cite{alpha_loss} per cycle for both the outer loop ($\alpha_{out}$), the QZI loop ($\alpha_{QZI}$) and initial and final optics ($\alpha_{init,final}$).  Hence the total transmission efficiency of the OAM spectrometer becomes $P(l_0) \approx \alpha^{(2N+2)(l_0+1)}\cos\left(\frac{\pi}{2N}\right)^{2Nl_0}$ for the $l_0$-th order OAM component.  We plot in Fig.~\ref{fig:loop_opt_perfectFilter}(b) the $P(l_0)$ vs. $N$ for OAM components $l_0=0-10$.  With increasing $N$, the quantum Zeno effect leads to an increase in $P(l_0)$, while loss leads to a decrease in $P(l_0)$.  As a result, an optimal $N$ is found at about $7-8$ for high order OAM components.  Note that the extinction ratio between different OAM states remains infinite even in the presence of loss. Crosstalk would only take place when the OAM filter is not completely opaque to nonzero OAM states.

To take into account imperfect OAM filters, we derive below the general expression for the transmission efficiency and extinction ratio, with finite $N$ and optical loss.  We consider the OAM filter having a  complex transmission coefficient $\sqrt{T(l)}e^{i\phi(l)}$ for the $l$th OAM component.  If the state $|\psi\rangle =|H,l_0\rangle$ enters the QZI, after $N$ cycles, it exits the QZI loop in a polarization superposition state $p_H|H\rangle + p_V|V\rangle$ \cite{jang_optical_1999}, where $p_H$ and $p_V$ are given by:

\begin{equation}
\left( \begin{array}{c}
p_H \\
p_V \\\end{array} \right)
 = \alpha_{QZI}^{N}\left[
\left( \begin{array}{cc}
1 & 0 \\
0 & \sqrt{T(l)}e^{i\phi(l)} \\\end{array} \right)
\left( \begin{array}{cc}
\cos\left(\frac{\pi}{2N}\right) & \sin\left(\frac{\pi}{2N}\right) \\
\sin\left(\frac{\pi}{2N}\right) & -\cos\left(\frac{\pi}{2N}\right) \\\end{array} \right)
\right]^N
\left( \begin{array}{c}
1 \\
0 \\\end{array} \right).
\label{eq:QZI_final_state}
\end{equation}
The pulse re-enters the outer loop at PBS3 with probability $|p_H|^2$, corresponding to a successful interrogation by the QZI (if $l_0 \neq 0$).  With probably $|p_V|^2$, the pulse exits toward the detector, corresponding to an error (if $l_0 \neq 0$).  The total loss of this QZI interrogation is $|loss|^2 = 1-|p_H|^2-|p_V|^2$.  In the outer loop, the OAM value of the pulse is lowered by $1$ via the VPP, and the intensity of the pulse is reduced by a factor $|\alpha_{out}|^2$ per loop.  Therefore, the probability of detecting the OAM eigenstate $l_0$ in the $l$th time interval (or, measured as with OAM $l\hbar$) is given by:
  \begin{equation}
  P(l;l_0) = |\alpha_{init,final}|^2|p_V(l_0-l)|^2 \prod_{m=l_0-l+1}^{l_0}\left(|\alpha_{out}|^2 |p_H(m)|^2\right).
  \label{eq:PPll0}
  \end{equation}
And we define the extinction ratio $\eta$ as:
\begin{equation}
\eta(l_0) = P(l_0;l_0)/\sum_{l\neq l_0} P(l;l_0).
\end{equation}

With an imperfect OAM filter, light with nonzero OAM has a finite probability of transmitting through the filter in vertical polarization after the $N$th QZI-loop.  It will then be switched to horizontal polarization by P1 and exit at a time interval corresponding to components with a lower OAM. Consequently, the extinction ratio is reduced.  If the light is transmitted through the filter before the $N$th loop, it will results in a larger loss.  An imperfect OAM filter may also partially block light with zero OAM, which which also results in loss.

\begin{figure}[tb]
\centering
    \includegraphics[width=.98\textwidth]{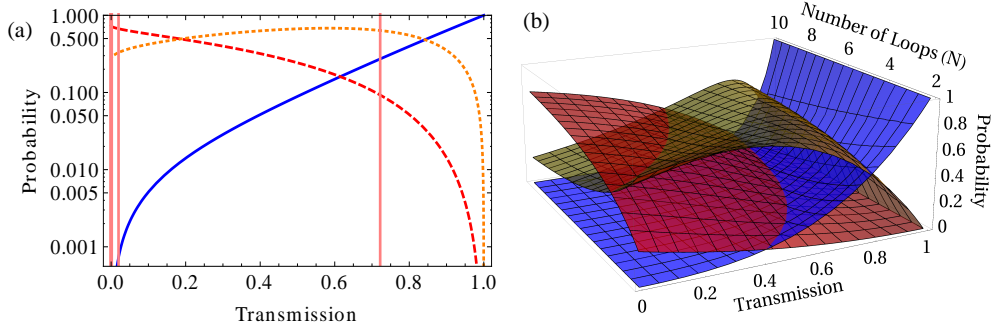}    
  \caption{The probabilities of different outcomes of a QZI interrogation as a function of the transmission of the OAM filter, neglecting optical loss. The blue solid line represents detecting OAM=0, the red dashed line is detecting OAM$\neq 0$, and the orange dotted line, loss. (a) $N = 8$. (b) $N=2-10$.}
  \label{fig:qzi_outcomes}
\end{figure}

Figure~\ref{fig:qzi_outcomes}(a) shows, per quantum Zeno interrogation of light with OAM of $l\hbar$, the probabilities of the light exiting toward the detector ($|p_V|^2$), re-entering the outer-loop ($|p_H|^2$) and being lost ($1-|p_V|^2-|p_H|^2$).  These probabilities are plotted as a function of transmission $T(l,a_0)$ and $N$.  The crossing of $|p_H|^2$ and $|p_V|^2$ separates the regimes when the interrogation result is more likely (to the right side) or less likely (to the left side) to be correct than incorrect.

\begin{figure}[tb]
  \centering
    \includegraphics[width=.98\textwidth]{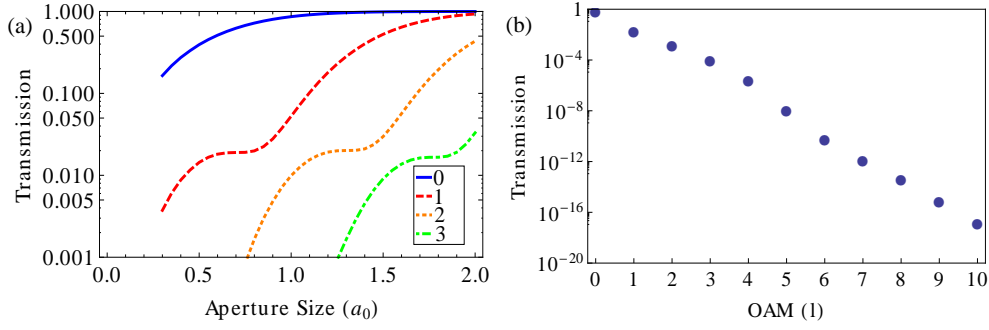}      
  \caption{Transmission of the pinhole spatial filter (a) as a function of the normalized aperture size $a_0$, for OAM components with $l_0=0-3$ and (b) as a function of $l_0$ with $a_0= 0.8$.}
  \label{fig:oam_filter}
\end{figure}
\begin{figure}[htb]
  \centering
    \includegraphics[width=.98\textwidth]{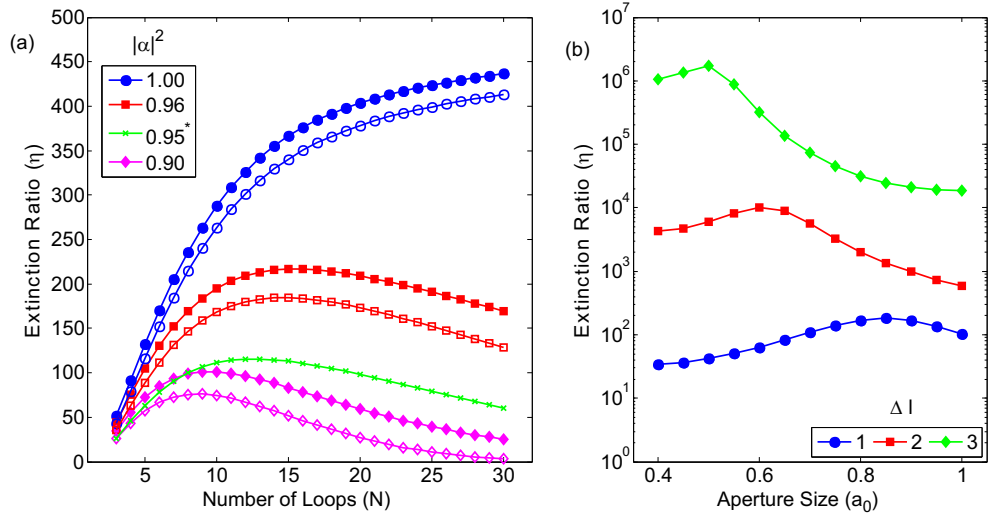}      
  \caption{(a) Extinction ratio $\eta$ as a function of the number of loops $N$ for various losses $|\alpha|^2$.  Solid symbols are for $l_0=1$ and open symbols are for $l_0=3$. $l_0> 3$ are essentially indistinguishable from $l_0=3$.  For the $l_0=0$ case, the extinction ratio is over a 1000 for all $|\alpha|^2$ values because no premature measurements are possible.  The additional green crosses labeled as $|\alpha|^2=0.95^*$ represents $|\alpha|^2=0.96$ but including misalignment of the OAM filter and VPP as discussed in the text.  (b) Extinction ratio $\eta$ as a function of the normalized aperture size $a_0$ for $l_0=6$, $\Delta l=1-3$,  $N=8$, and $|\alpha|^2 = 0.96$.  Skipping OAM states increases the extinction ratio by orders of magnitude. }
  \label{fig:loop_opt}
\end{figure}

As a practical example of an imperfect OAM filter, we consider a pinhole spatial filter.  Light with OAM of $l\hbar\neq 0$ has zero intensity at the center of the beam, while light without OAM has maximum intensity at the center.  Hence a very simple pinhole efficiently distinguishes light with and without OAM.  The intensity distribution of a Laguerre-Gaussian beam, a paraxial beam possessing OAM $l\hbar$, is given by \cite{allen_orbital_1992}:
\begin{equation}
I_{LG}(l;\rho ) = \frac{I_0}{\int_0^\infty du u^{|l|} e^{-u} L_{|l|}(u)} \left(\frac{\sqrt{2}\rho}{w_0}\right)^{|l|} L_{|l|}\left(\frac{2\rho}{w_0^2}\right) e^{-\frac{\rho^2}{w_0^2}}
\label{eq:lg_intensity}
\end{equation}
Where $L_l(x)$ is the $l$th order Laguerre Polynomial.  Thus the transmission $T(l)$ through a pinhole with a radius $a_0$ (normalized by the waist of the $l_0=0$ Gaussian beam) is:
\begin{equation}
T(l,a_0) = \left.{\int_0^{a_0}\int_0^{2\pi} \rho d\rho d\phi I_{LG}(l;\rho )}  \right/  {\int_0^\infty \int_0^{2\pi} \rho d\rho d\phi I_{LG}(l;\rho )}
\label{eq:pinhole_trans}
\end{equation}

Figure~\ref{fig:oam_filter}(a) shows $T(l,a_0)$ vs. $a_0$ for $l=0-3$. The transmission decreases sharply with increasing $l$ when $a_0$ is smaller than $\sim 0.8$.  Choosing $a_0=0.8$, we show in Fig.~\ref{fig:oam_filter}(b) the nearly exponential decrease of $T(l,a_0)$ with $l$.  These values are also marked by the red vertical lines in Fig.~\ref{fig:qzi_outcomes}(a).  Due to the fast decrease of $T(l, a_0)$ from $l=0$ to $l\geq 1$, a large extinction ratio is readily achieved, which is very well approximated by:
\begin{equation}
\eta(l_0) = P(l_0;l_0)/\sum P(l\neq l_0;l_0) \approx \frac{\alpha_{out}|p_V(0)|^2 |p_H(1)|^2}{|p_V(1)|^2}.
\label{eq:approx_extinction_ratio}
\end{equation}
$\eta$ is essentially the same for all OAM components, and it is mainly determined by how well the QZI can distinguish between states with OAM values $l_0=0$ and $l_0=1$.  We plot in Fig.~\ref{fig:loop_opt}(a) $\eta$ vs. $N$ for $|\alpha|^2=0.9-1$.  In general, $\eta$ increases with $N$ but decreases with $|\alpha|^2$, resulting in an optimal $N$ for each $|\alpha|^2<1$.  Even for $|\alpha|^2=0.9$, $\eta>70$ can be reached with $N=7$.  For $|\alpha|^2=0.96$, $\eta$ peaks at $\sim 180$.

\begin{figure}[tb]
  \centering
    \includegraphics[width=.95\textwidth]{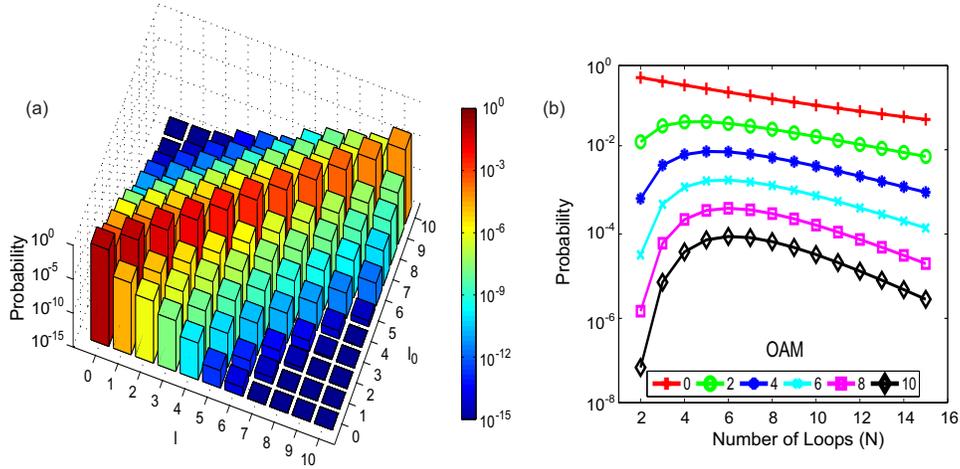}      
  \caption{(a)The probability of measuring an OAM value $l$ for a given input state $l_0$ (Equation~\ref{eq:PPll0}), using pinhole as the OAM filter, $N=8$, $|\alpha|^2=0.96$, and misalignment of 10\% and 1\%, respectively, at the pinhole filter and VPP. Despite the decrease in probability for the diagonal elements at large $l_0$, the off diagonal elements decrease much faster, as implied by the large extinction ratios. (b) The diagonal elements of (a) as a function of $N$ for $l_0=0-10$.  }
  \label{fig:abs_prob_real}
\end{figure}

  An additional source of error is due to the misalignment of the beam through two OAM-sensitive components: the OAM filter (e.g. a pinhole) and the VPP. Misalignment at the pinhole filter leads to reduced coupling efficiency of the zero OAM state, increased transmission of non-zero OAM states, and thus reduced extinction ratio. Misalignment on the VPP changes the desired OAM state into a superposition with neighboring OAM orders. However, these neighboring orders have very small amplitudes (e.g. $<1\%$ with $1\%$ misalignment) \cite{molina-terriza_management_2001}, and they are further filtered out through the QZI loop, resulting in negligible reduction in the extinction ratio. The main effect of misalignment at VPP is the slightly reduced transmission of the correct OAM state, hence reduced overall detection probability. We illustrate the effects of misalignment on the extinction ratio in Fig.~\ref{fig:loop_opt}(a) (the green crosses), assuming conservatively $10\%$ misalignment of the focused beam waist at the pinhole and $1\%$ misalignment of the collimated beam waist at the VPP. Extinction ratios over 100 are still readily achieved.

  The extinction ratio can be increased by many orders of magnitude if we only need to measure every other order, or every third order of OAM (Figure~\ref{fig:loop_opt}(b)).  Correspondingly, we can choose smaller aperture sizes and a VPP that reduces the $l$ by $\Delta l=2$ or $3$ per passing.  A smaller aperture size also introduces extra loss, but only in the final QZI on the zero OAM state, and thus only decreases the detection probability by about a factor of two.

To evaluate the overall performance of the OAM spectrometer, we plot in Fig.~\ref{fig:abs_prob_real}(a) $P(l;l_0)$ vs. $l$ and $l_0$ on the log scale, including loss and misalignment.  The diagonal elements $P(l_0;l_0)$ correspond to correctly detecting an OAM component. They are two orders of magnitude higher than neighboring off diagonal elements, consistent with the high extinction ratios calculated before. In Fig.~\ref{fig:abs_prob_real}(b), we show $P(l_0;l_0)$ as a function of $N$ for different $l_0$. $N \sim 8$ gives the highest probability for detecting high order OAM components, while still maintaining an extinction ratio of above 100.

  In summary, we present a compact OAM spectrometer that disperses light of different OAM values in time.  Loss is significant for high order OAM components with commercially available optical components.  However, the high loss doesn't have an appreciable effect on the signal to noise ratio; extinction ratios of $>\!100$ are readily achieved even after taking into account optical loss and misalignments. The extinction ratio can be further improved by many orders of magnitude by skipping OAM orders, or by using a better OAM filter than a simple pinhole.


\begin{thebibliography}{99}
\bibitem{alpha_loss}
Technically, each $|\alpha|^2$ is slightly different, but the difference is well within 1\%. The $|\alpha|^2$ that matters most is the one corresponding to the QZI loop (including S1), which, without any optimization, consists of 4 beam splitters, 5 mirrors, 1 waveplate and up to three Pockels cells. Assuming all optics are anti-reflection coated so that loss is 1\% at each Pockels cell and 0.1\% at each other component, we have $|\alpha|^2\geq 0.96$.
\end{thebibliography}
\end{document}